# Modeling the variability of shapes of a human placenta.
**December 20, 2007**

Michael Yampolsky[1], Carolyn M. Salafia[2], Oleksandr Shlakhter, Danielle Haas, Barbara Eucker, John Thorp

While it is well-understood what a normal human placenta should look like, a deviation from the norm can take many possible shapes. In this paper we propose a mechanism for this variability based on the change in the structure of the vascular tree.

> Happy families are all alike; every unhappy family is unhappy in its own way.
>
> Leo Tolstoy, *Anna Karenina*

## Introduction.

The placenta is the sole fetal source of oxygen and nutrients. As such, it is a principal regulator of fetal growth and fetal health. Normal placentas will grow uniformly out from the umbilical cord insertion, resulting in a round to oval disk with a centrally inserted cord. A variable maternal uteroplacental environment (the maternal "soil") affects macroscopic placental structure as a change in shape. Where the maternal "soil" is not receptive, placentas will not grow, or not robustly. Irregularities in disk outline, umbilical cord insertion (*Figure 1a-b*) and in disk thickness (*Figure 2a-b*) are markers of fetal-placental environmental pathology, denoting variable placental arborization, and as such, deformation of normal placental

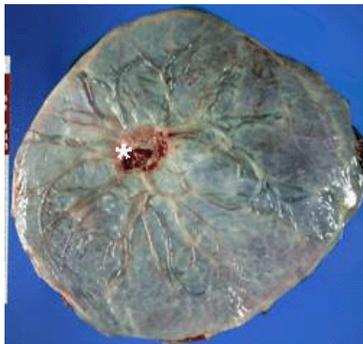

**Figure 1a.** *Normal round-oval placental disk, * marks umbilical cord insertion.*

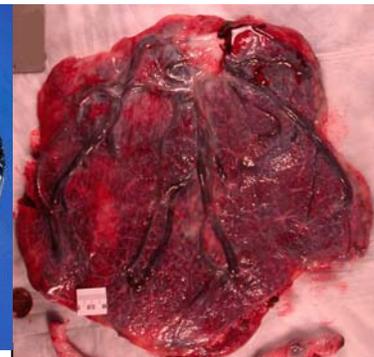

**Figure 1b.** *Irregularly scalloped disk, cord insertion top center.*

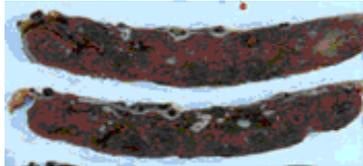

**Figure 2a.** *Uniform disk thickness reflecting uniform arborization of chorionic villi.*

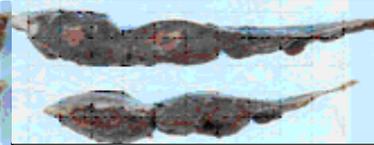

**Figure 2b.** *Variable disk thickness, variable placental growth suggests variable maternal uteroplacental "soil".*

[1] M.Y. was partially supported by NSERC Discovery Grant
[2] C.S. was partially supported by NARSAD Young Investigator Award, K23 MidCareer Development Award and NIMH K23MH06785

growth resulting in an abnormal placental structure. The microscopic growth of the human placenta involves repeated branching, analogous to the roots of a tree; its mature arborization pattern is complex (e.g., **[2-11]**, *Figures 3 and 10*), so complex that it cannot be measured reliably even by expert, dedicated pediatric pathologists **[12, 13]**. Just as the pattern of roots reflects the underlying soil's fertility and predicts the health of plants that depend on those roots for sustenance, placental arborization reflects the health of the maternal environment and impacts on fetal health **[14]**.

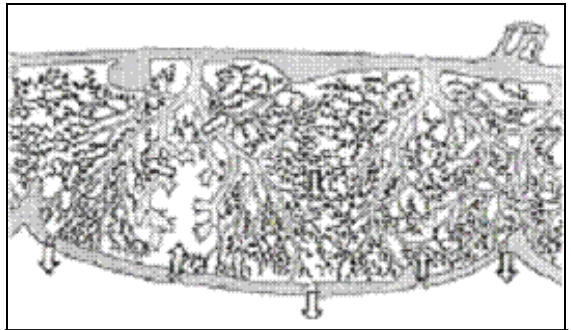

**Figure 3.** *This placental cross-section shows the complexity of placental branching. The arrows show the maternal blood flow that percolates between villous branches; nutrient and oxygen exchange occurs in the distal placental branch tips.*

The normal shape of a human placenta is well-understood **[15]**, however, there are many possible deviations from the norm. Shapes "other than round" can be difficult to classify; the Collaborative Perinatal Project used a variety of terms to attempt to describe such abnormal placental shapes but had to resort, after "bipartita" and "tripartite" to terms such as "multiplex" to convey the complexity of placental shapes. Such subjective and imprecise terminology has not advanced our understanding of the genesis of such shapes, and has limited our ability to analyze the relationship of abnormal placental growth shapes to the health of the fetus and child.

However, it is evident, that many abnormally shaped placentas are quite regular, and can be classified into several well-defined geometrical patterns. This regularity suggests that there may be a common underlying pathological mechanism(s) responsible for much of the variability of observed shapes of placentas. In this work we propose such a mechanism: we derive the variability of placentas from a change in the branching structure of their vascular trees. To test this hypothesis, we introduce a dynamic model for the growth of the vascular tree of a placenta. This model is based on a biologically realistic random growth process. We then show how a change of the parameters of the growth at a single instance leads to the appearance of the observed variability of placentas.

**Acknowledgment.** This project was started during 2007 Program "Random Shapes" at the Institute for Pure and Applied Mathematics at UCLA, where C.S. and M. Y. were Core participants. We gratefully acknowledge the support of IPAM, and thank the organizers of the program for bringing us together. M.Y. would like to thank two other participants, Ilia Gruzberg and Ilia Binder for useful discussions of DLA.

**Materials and Methods**

*Placental Cohort.*
The *Pregnancy, Infection, and Nutrition Study* is a cohort study of pregnant women recruited at mid pregnancy from an academic health center in central North Carolina. Our study population and recruitment techniques are described in detail elsewhere (**[58]**). Beginning in March 2002, all women recruited into the *Pregnancy, Infection, and Nutrition Study* were requested to consent to a detailed placental examination. As of October 1, 2004, 1,159 women (94.6 percent) consented to such examination. Of those women who consented, 1,014 (87.4 percent) had placentas collected and photographed for image analysis. Of the 1014 consecutive placentas collected, two cases were excluded because the trimmed placental weight was not recorded, and six cases were delivered in fragments, such that measurement of chorionic plate landmarks was not possible. This left 1008 cases for analysis, 99 percent of the available placental sample.

Placental gross examinations, histology review, and image analyses were performed at *EarlyPath Clinical and Research Diagnostics*, a New York State-licensed histopathology facility under the direct supervision of Dr. Salafia.
The institutional review board from the University of North Carolina at Chapel Hill approved this protocol.

The fetal surface of the placenta was wiped dry and placed on a clean surface after which the extraplacental membranes and umbilical cord were trimmed from the placenta. The fetal surface was photographed with the Lab ID number and 3 cm. of a plastic ruler in the field of view using a standard high-resolution digital camera (minimum image size 2.3 megapixels). A trained observer (D.H.) captured series of *x,y* coordinates that marked the site of the umbilical cord insertion, the perimeter of the fetal surface, and the "vascular end points", the sites at which the chorionic vessels disappeared from the fetal surface. The perimeter coordinates were captured at intervals of no more than 1 cm, and more coordinates were captured if it appeared essential to accurately capturing the shape of the fetal surface. The chorionic vessels extend out from the umbilical cord insertion and, at varying intervals from the edge of the fetal surface, dive beneath the surface so that they are no longer visible. Those "vascular end points" mark the terminal differentiation of the chorionic vessels into fetal stems and the finer structures of the placental functional units (P. Kaufman, MD, personal communication).

**Seeing Stars (I).**

The normal shape of a placenta is round, with the umbilical cord insertion roughly at the center, as illustrated in the figure below.

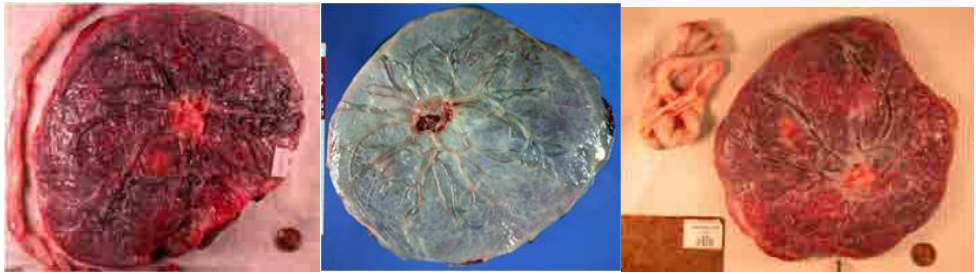

**Figure 4:** *normal placentas.*

We aim to explain some of the variability from the normal, concentrating on two regular shapes which appear prevalent:

- Star-shaped placentas (about 5% of all cases). A typical star has between 5 and 7 prominent spikes. Variability of the radius of a star, as measured from the umbilical cord, is typically within $20 - 30\%$.

**Figure 5:** *stars.*

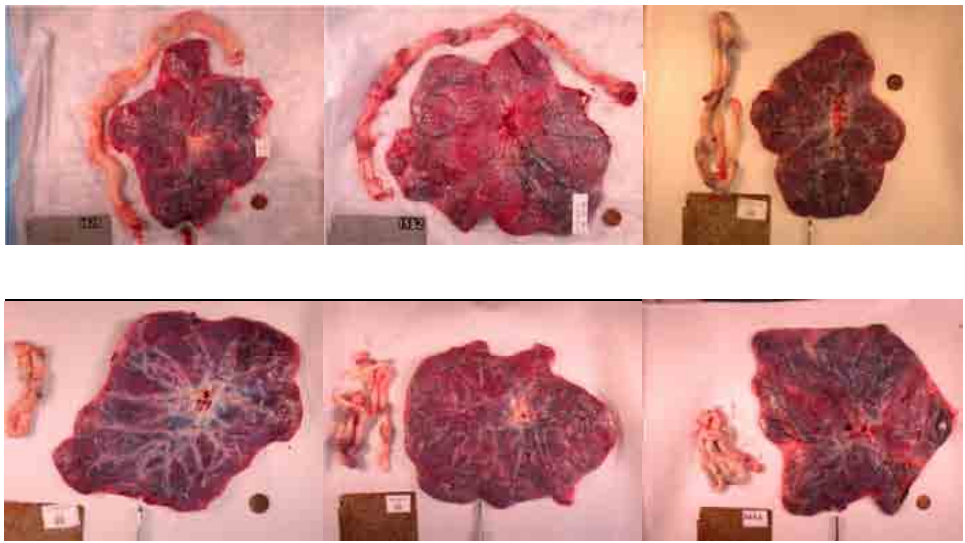

- Several pronounced lobes. These placentas also account for about 5% of the total cases. The number of lobes is typically 2 or 3.

**Figure 6:** *placentas with several lobes.*

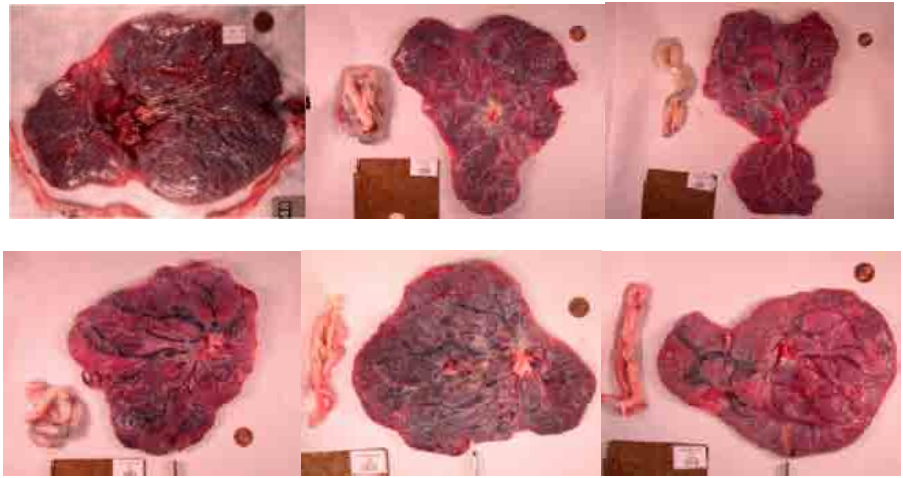

We propose to explain this observed variability by a deviation in the growth of the vascular tree of the placenta. In the next section we introduce a dynamic growth model for a vascular network. We then demonstrate how the model explains the observed shapes, and present evidence to support our explanation.

**A note on the existing models of vascular trees.**

Numerous *static* models of vascular trees exist in the literature (see e.g. **[16], [17]**)**.** They are usually based on various optimization algorithms for filling the spatial structure of an organ. Such an approach is, however, completely unsuitable for our purposes. We aim to demonstrate how a deviation in the process of the *dynamical* growth of the vascular network affects the macroscopic shape of a placenta. We thus present a *dynamic* model of growth of a vascular tree.

**Modeling growth at the tips, or a cloud of blind flies.**

Growth of blood vessels is recognized to depend on the concentration gradients of appropriate growth factors. Cells situated at the tip of vascular sprouts sense and navigate the environment, while cells in the sprout stalks proliferate and form a vascular lumen. Migration of the tip cells depends on a graded distribution of VEGF-A and activation of VEGFR2 located on the tip-cells. Proliferation in the stalk is concomitantly regulated by the local VEGF-A levels. Thus, the shape of the VEGF-A gradient controls the balance between tip cell migration and stalk cell proliferation, which in turn determines the initial vascular pattern **[48].** VEGF is an important regulator of placental angiogenesis **[49], [50]**, although the specific cell physiology has not been as well studied.

How do we model the new growth of a vascular tree at the tips of the branches? For simplicity, let us discretize both time and the units of growth. For a three-dimensional tree we should then ask where its tips are, to determine where the growth is likely to occur in the next unit of time. While this does not appear to be a simple task even for a tree with a few branches, there is a general (and easily automated) method for doing this.

For illustration, consider a tree. At some distance from the tree, release a cloud of blind flies. Flying randomly, without any sense of direction, our flies will only stop when they come in contact with the tree. At this point, a fly would close its wings, and sit at the spot where the contact occurred. It is a remarkable mathematical fact that:

*It is one of the tips of the tree, where a fly is most likely to land.*

We can thus identify the tips as the places with the most flies, and the relative number of flies which have landed at a given extremity will correspond to the likelihood that the next unit of growth will happen there.

> **Figure 7:** a two-dimensional illustration of this effect. A tree with eight symmetric branches, and the positions where the first 100 flies land on it, generated by a numerical simulation (marked with dots).

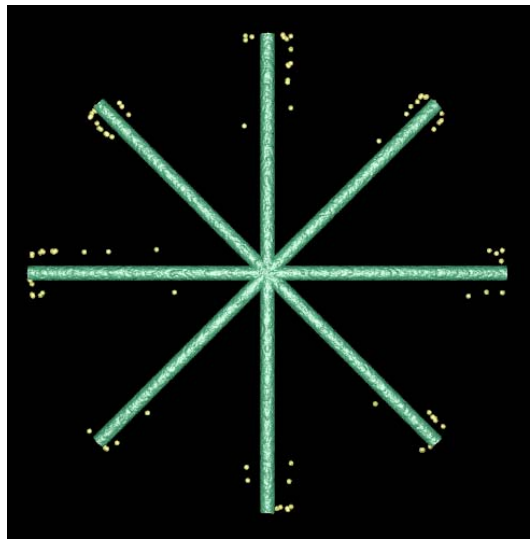

To formulate this in more precise terms, let us select as a unit of growth (which we will call a *particle*), a three-dimensional ball of a small diameter $d$. At each moment of time our tree grows by one particle. At the initial time $t=1$, we start with a tree $C_1$ consisting of a single particle. Thus, at time $n$, our tree is a *cluster* $C_n$ consisting of $n$ particles. To grow it by one unit, consider a large sphere $S$ around $C_n$ (much larger than the size of the cluster itself). Randomly select a point $w$ in $S$, as the initial position of a blind fly. The fly will perform a *random walk* (or a random flight in this case) originating from $w$. That is, it will randomly perform a sequence of moves up/down, left/right, forward/backward. Recording its spatial positions, we will see a sequence of points

$$w_0=w, \ w_1,\ldots, \ w_{t-1}, \ w_t,\ldots$$

where $w_t$ is produced from $w_{t-1}$ by adding an increment $s_t$. The increment always has the same size, comparable to *d* (to fix the ideas, we can take it equal to *d*). The direction of the increment has to be aligned with one of the three coordinate axes (up/down, left/right, forward/backward) and randomly chosen from the six possibilities.

If at some point of time *t* the point $w_t$ is at a distance of less than *2d* from the cluster $C_n$, then we attach a new particle to this position, and thus obtain a bigger cluster $C_{n+1}$. If before this happens, the particle $w_t$ drifts outside of the large sphere *2S*, then we discard it, and select a new initial point *w*.

Such random growth, first introduced by Witten and Sander in 1981 (**[18]**, **[19]**), is known to physicists under the somewhat unwieldy name *Diffusion Limited Aggregation*, or *DLA* for short.

> Mathematically, the DLA growth is described as follows. Consider the electrostatic potential of the cluster $C_n$, that is, the solution of the Laplace equation
>
> $$\Delta\Phi=0$$
>
> with the boundary condition $\Phi=0$, and with the asymptotics $\Phi(z) \sim -1/z$ at infinity. Now the probability of the new growth at a boundary point of $C_n$ in a given direction is proportional to the magnitude of the gradient $\nabla\Phi$ (the strength of the electric field).

**Visualizing a DLA tree as a vascular network.**

A tree $C_n$ constructed in this way consists of *n* small three-dimensional balls. To visualize it consistently with the structure of a vascular network, we first identify its branch structure in the following way. We say that a particle *x* is an immediate descendant of a particle *y*, if *x* is attached to *y*, and *y* is older than *x*. Further, $x_n$ is a descendant of $x=x_0$ if there is a chain of DLA particles

$$x_0, \ x_1,\ldots, \ x_{n-1}, \ x_n$$

such that $x_i$ is an immediate descendant of $x_{i-1}$ for $0 < i \leq n-1$. Now we can assign a visualization size *v* to each DLA particle, consistently with the principle, that bigger vessels grow first. Specifically, let the weight *m* of a particle *x* be the total number of its descendants. Thus, older particles, which are located closer to the root (the very first particle $C_1$) of the DLA tree, will have a larger weight. We then assign

$$v = \text{📌}^{\text{📌}}.$$

**Growing denser trees.**

Our model has a single parameter $0<\kappa\leq 1$ which will be completely crucial to our study. It is the probability that a DLA particle will stick to the cluster when it collides with it. Think again of a blind fly which has encountered one of the tips of the tree. If $\kappa \neq 1$, then with non-zero probability $1-\kappa$ the fly will continue its travels. Most likely, it will end up sitting at some nearby point of the tree, but not necessarily at the tip. The effect of this is to add some diffusion to DLA growth. The smaller the value of $\kappa$ is, the more "hairy" the resulting tree will become.

From modeling considerations, it is also useful to note that the physical size *d* (as opposed to the visualization size) of a DLA particle can be changed during the growth of the cluster. This is a quick and dirty way of changing the branching structure. A smaller particle will also tend to diffuse along the tips of the DLA cluster formed by larger particles, hence a reduction of *d* will create more branching.

**Growing a human placenta with DLA[3].**

To model the growth of a placental vascular tree, we surround our growth with a spherical constraint, modeling the walls of the uterus. The constraint absorbs any DLA particle which hits it from the interior, thus ensuring that no growth penetrates the constraint. The initial cluster marks the insertion of the umbilical cord. It is placed near the "south pole" of the constraint.

To ensure that the initial growth is directed towards the constraint (the umbilical cord does not grow away from the wall of the uterus), it is helpful to select the initial cluster $C_1$ consisting not of one, but of several DLA particles. The number of particles in $C_1$ is still insignificant, compared with the overall size of the DLA growth (in our experiments, it is between 0.01-1% of the total). We have experimented both with a deterministically defined $C_1$, and with a short initial cluster given by an opportune DLA growth, with identical results. The total number of DLA particles in a grown model vascular tree is between 150-200K.

Once the parameters, such as the position and the shape of the initial cluster have been chosen, the model was tested by varying the random number generator (RNG) used to grow DLA[4]. We have tested the model in two ways: by changing the "seed" of a given RNG (this changes the random number sequence but not the algorithm which produces it); and by changing the RNG itself. The results were very robust.

---

[3] See section "Computer software" for the description of programs used in the paper.
[4] See the list of random number generators used in the end of the paper.

**Figure 8:** *below are some examples of model vascular trees. On the left is an XY-plane projection, on the right is a YZ-projection. In the YZ-plane the spherical constraint is apparent.*

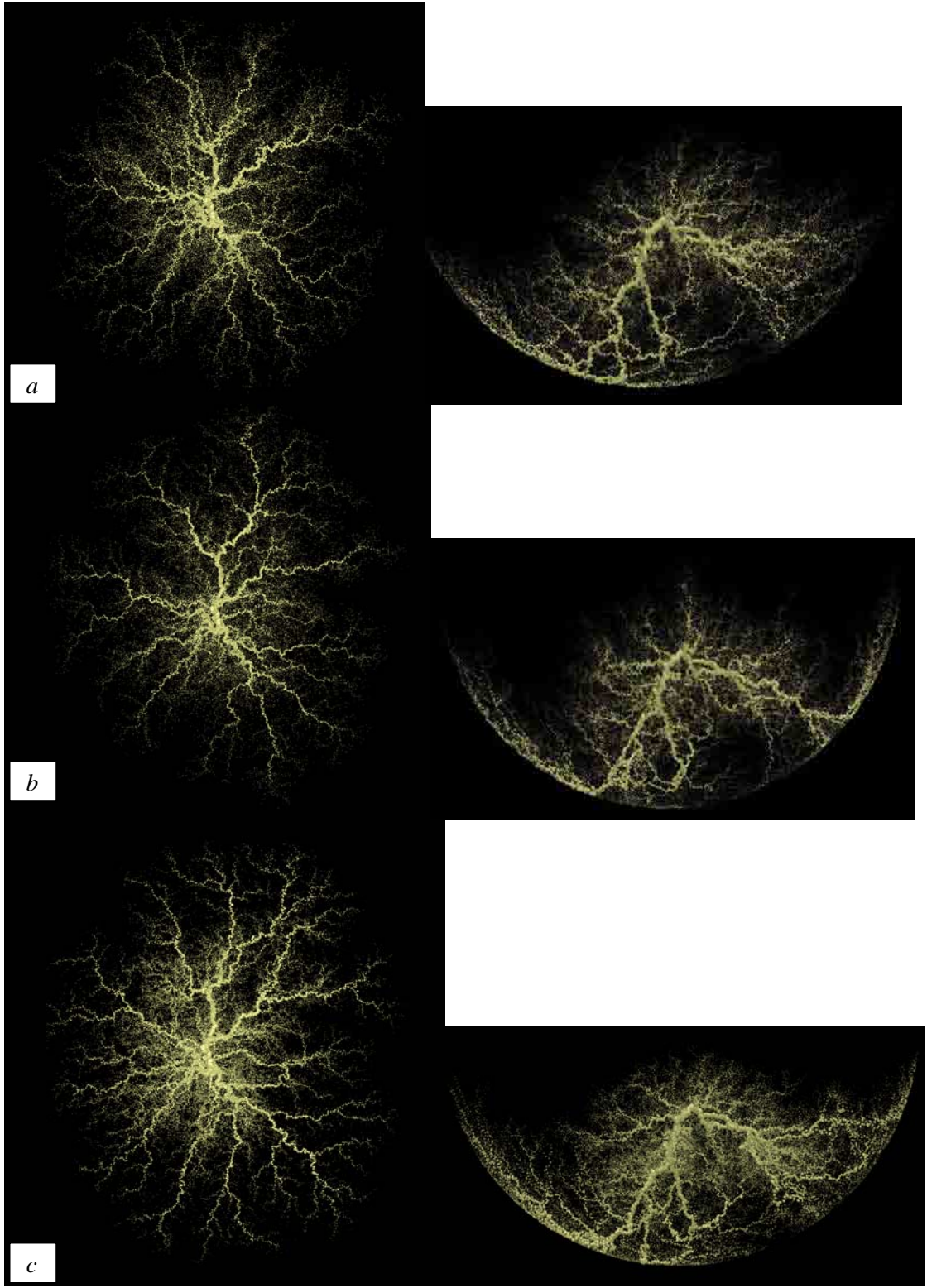

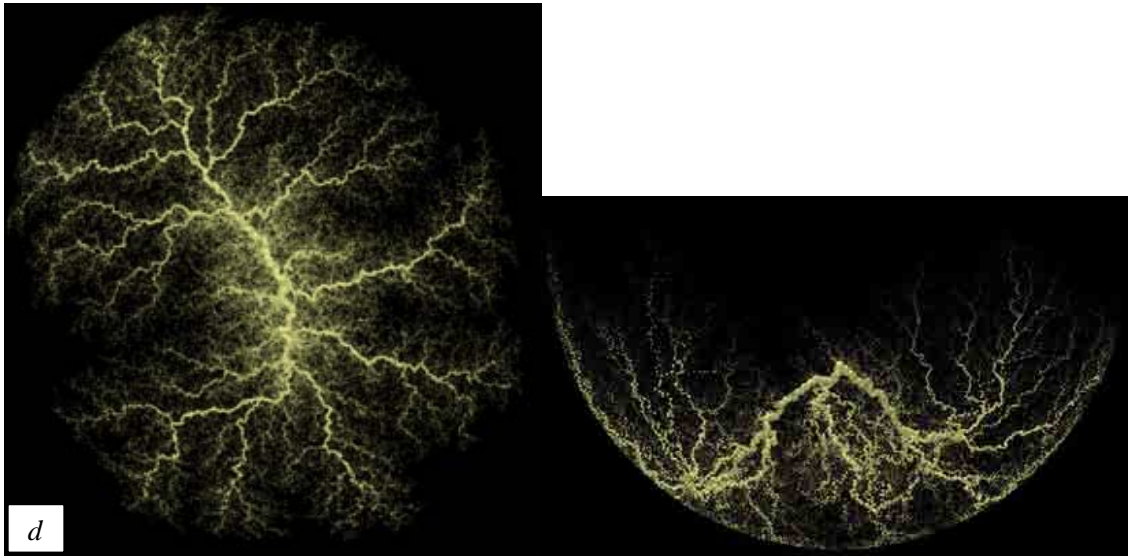

To visualize a model placenta from a vascular tree we can apply a "thickening" procedure: a white disk of a constant size is drawn in the background of each DLA particle. The resulting shape is then smoothed using a standard spline approximation. An example of this procedure applied to the first vascular tree from the previous figure is seen below.

**Figure 9:** *a thickened vascular tree models the macroscopic shape of a placenta.*

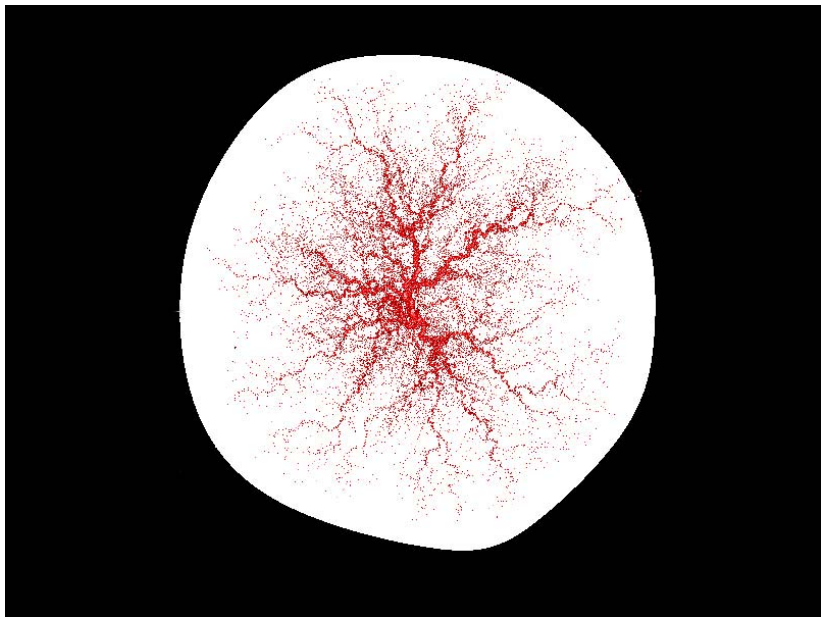

**Figure 10:** *for comparison, below are several X-Ray pictures of placentas injected with a radioactive dye (reproduced from [20]).*

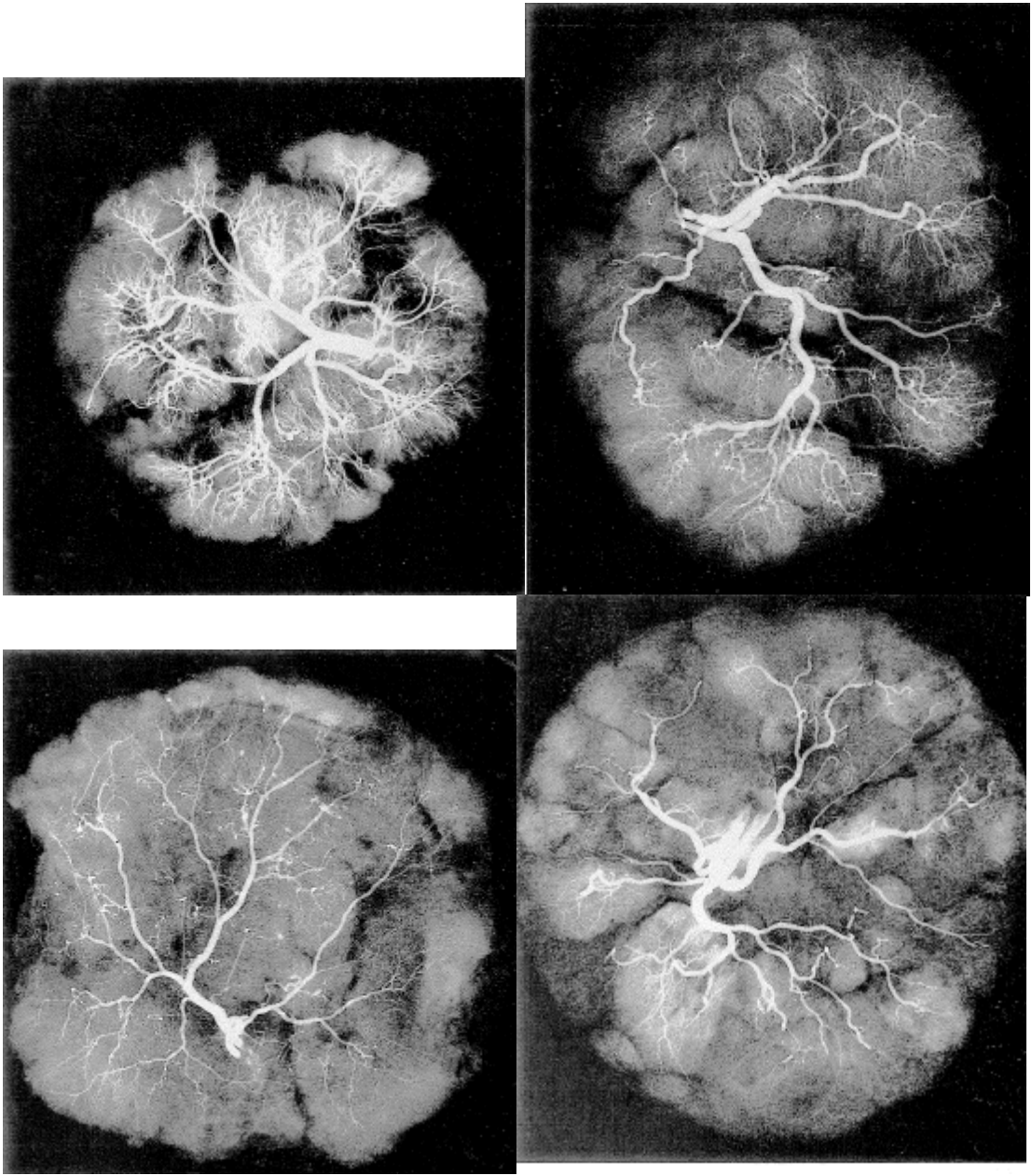

**Seeing Stars (II).**

To produce models of placentas of irregular shapes we have varied the branching parameter κ of the growth at various stages of the development of the model placenta. A single instance of the change in the value of κ over the course of growth of the model is required to produce the observed variability:

- Decreasing the branching parameter κ (by a multiple 0.01) at an early stage of the growth (after 5-7% of DLA particles have attached) leads to a picture with several (typically two or three) large lobes. Qualitatively, the large vessels grow apart early, and with an increased degree of branching, the tree does not produce enough "medium size" branches to uniformly fill a circular shape.
- Well-defined stars with 5-7 pronounced arms correspond to a change of branching at a late stage of growth (typically after 50% of DLA particles have attached). We have achieved this result by decreasing κ, and alternatively, by decreasing the physical size of the DLA particle. Changing the physical size of a DLA particle turns out to be a more convenient parameter here (the size was typically decreased by 20%).

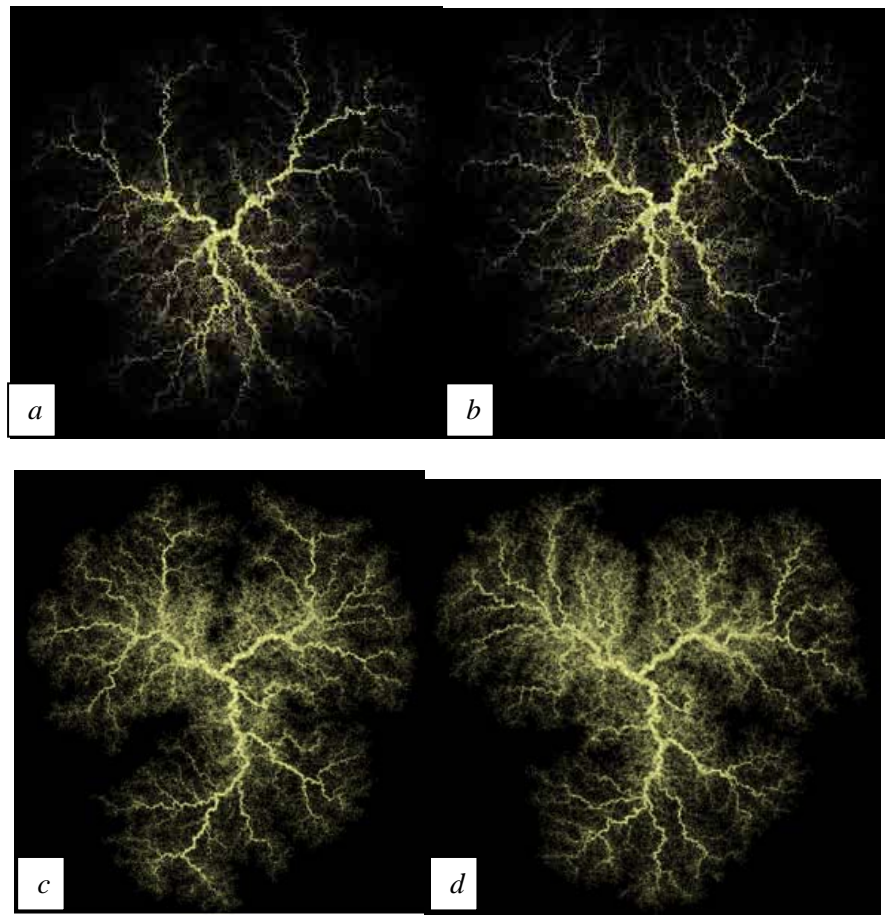

**Figure 11:** *model placentas with three large lobes. The lobes are more compact when the branching is increased at an earlier stage.*

**Figure 12:** *a placenta with three lobes (Fig. 11 a) visualized by adding white "shadows" to the blood vessels.*

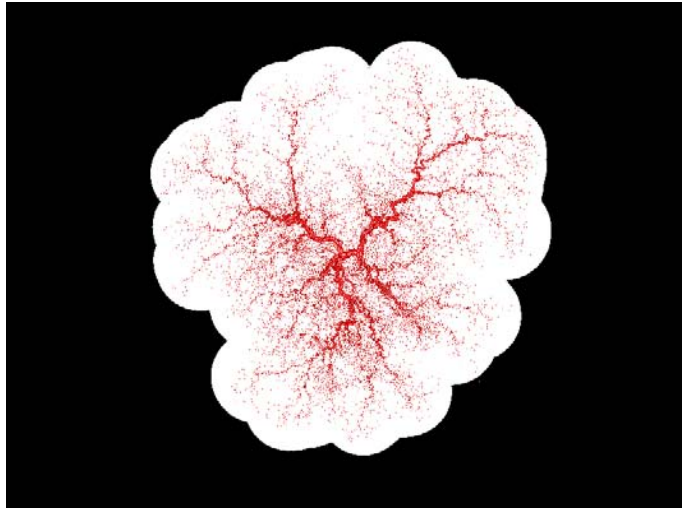

**Figure 13:** *For the same three-lobate placenta (Fig. 11a), the average number of small blood vessels is calculated in a sector of 60 degrees. A sector of this size is then rotated in increments of 10°, and the deviation from the average (as a percentage) is plotted in the chart below. The three lobes are clearly visible, together with the three gaps between them.*

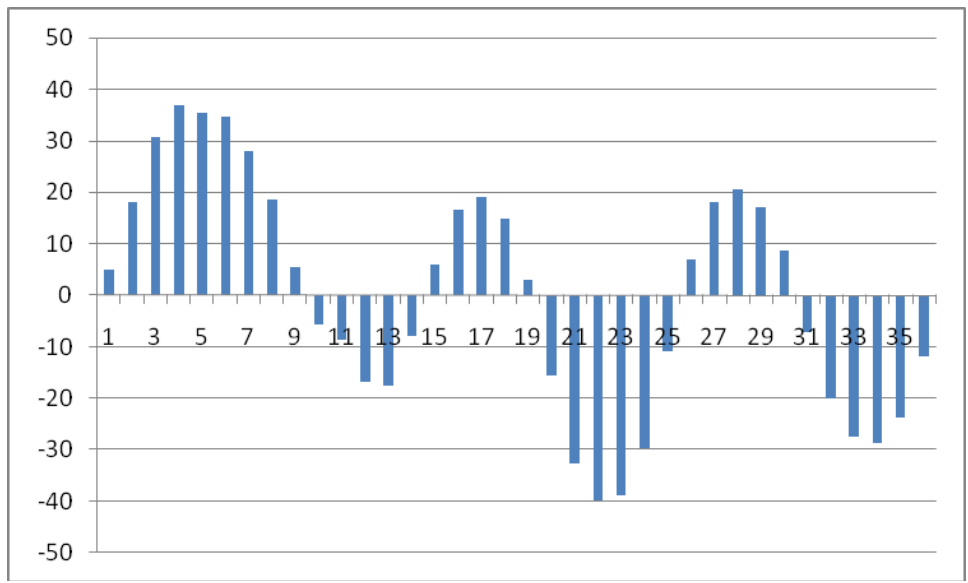

Below are several examples of model stars:

**Figure 14:** *star-shaped placentas:*

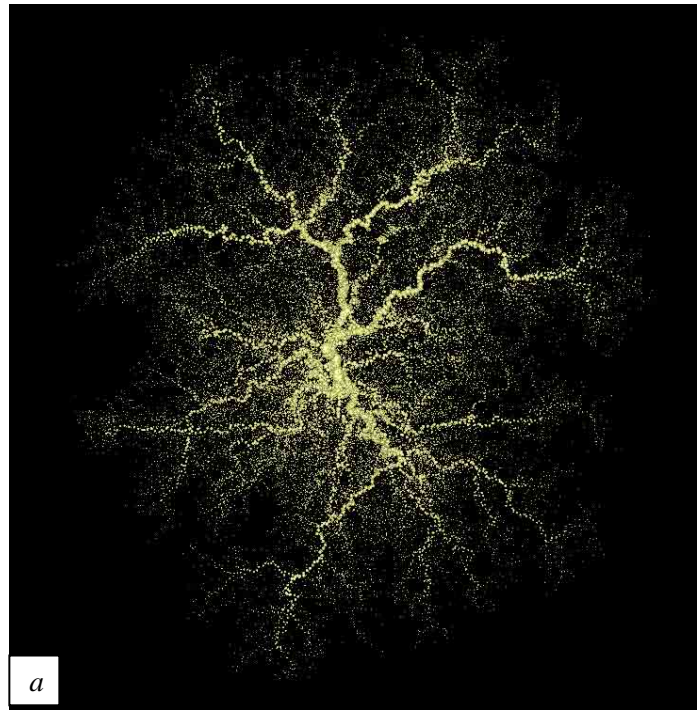

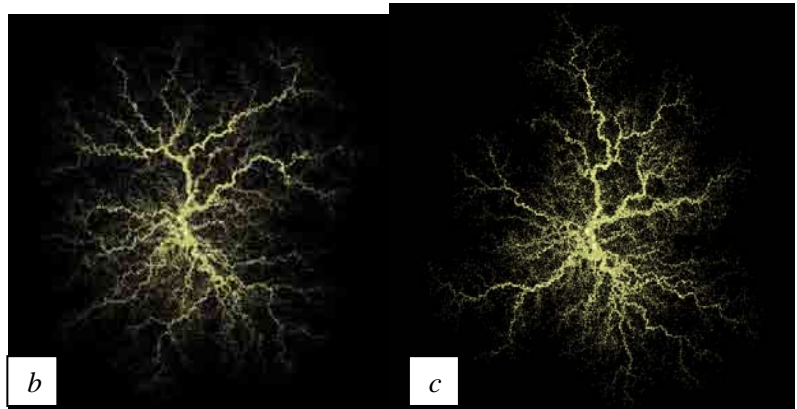

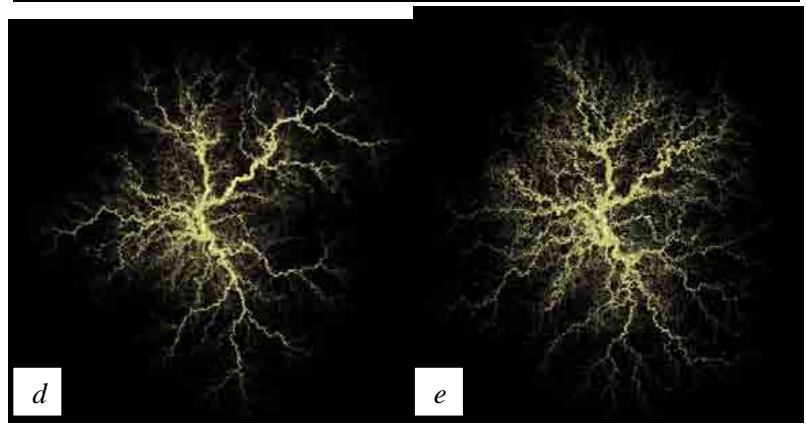

As further comparison of model shapes, we calculate the mean distance of a small vessel to the point of insertion of the umbilical cord in a sector of 5°, which is then rotated in increments of 2.5°. In each of the 5° sectors, we then mark the point on the bisector, whose distance to the vertex is equal to the mean distance for this particular sector. Connecting these points by line segments, we obtain a diagram, tracing the shape of the model placenta. Compare the results for a model a model round placenta (Fig. 8 c), a model star (Fig. 14 c), and a model three-lobate placenta (Fig. 11 a) in the figure below.

**Figure 15:** *the average distance of a blood vessel to the point of insertion of the umbilical cord in a model placenta (marked in sectors of 5°, rotated in increments of 2.5°. Left to right: a round placenta, a star, a three-lobate placenta.*

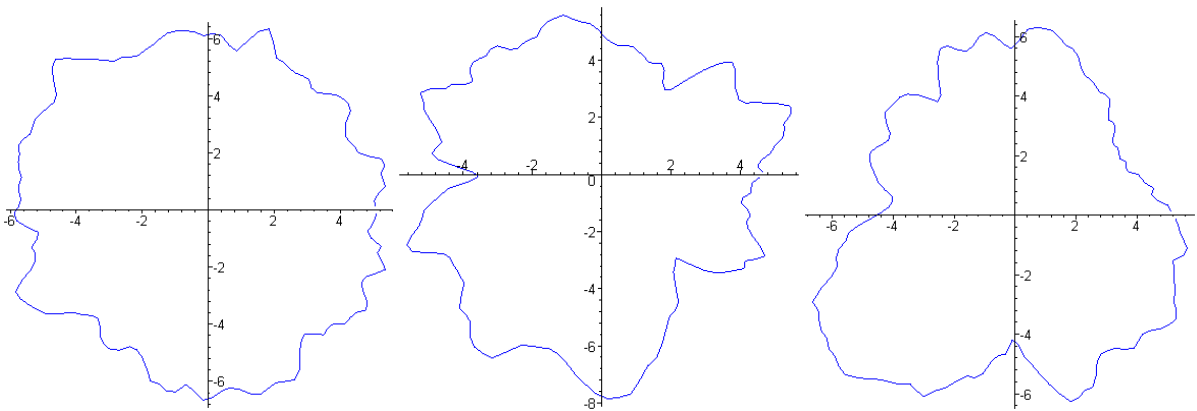

A further evidence for the mechanism of star formation we are describing is found in the variability of thickness of star-shaped placentas. A change in the density of the vascular tree is going to correspond to changing thickness of the placenta. Thus in a "star", the arms will be thicker than the areas between them. In a vertical slice, this would produce a wavy pattern, schematically shown below.

**Figure 16:** *a star on the left, with a position of a cut indicated. On the right is a vertical slice.*

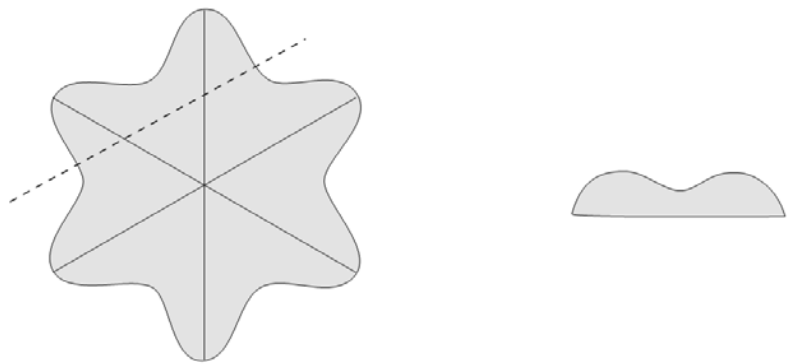

In the figures below, we demonstrate vertical slices of placentas taken at 1cm intervals. First is a typical normal placenta, with slices exhibiting uniform thickness. Next are pictures of slices of stars, with a wavy pattern clearly visible.

**Figure 17:** *slices of a regular placenta:*

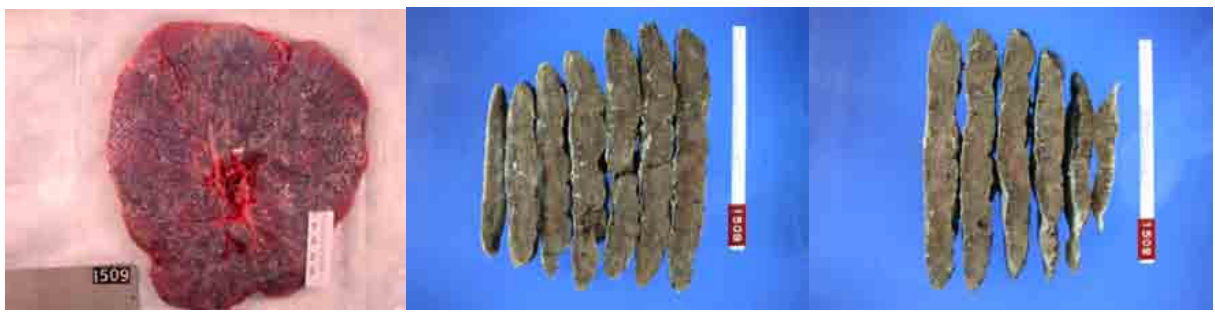

**Figure 17:** *slices of a star:*

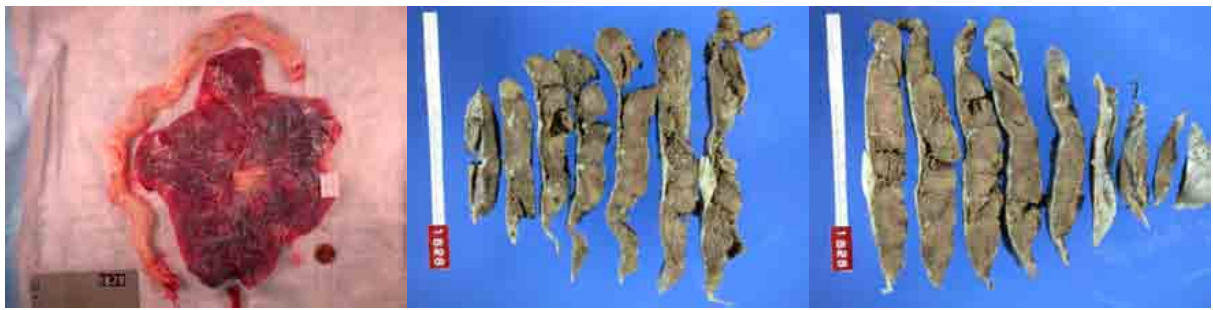

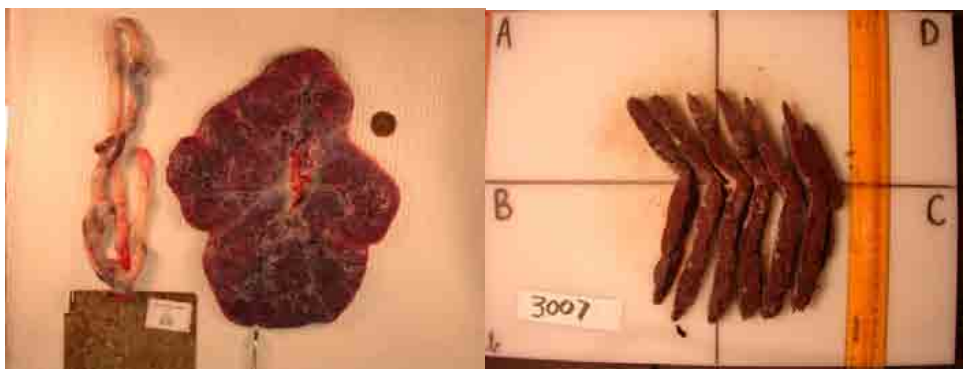

**Further evidence that variable shapes reflect the architecture of the vascular fractal: deviations from the ¾ rule.**

The *¾ rule* (or *Kleiber's Law*) is a famous allometric scaling law, which postulates that the basic metabolic rate *B* and the body mass *M* are related as

$$B \sim M^{3/4}.$$

If the vascular system consisted of thin tubes of small size, uniformly filling the volume of the body, one would expect the two variables to be proportional. The appearance of a power $3/4 \neq 1$ is explained by the fractal structure of the vascular tree (see e.g. **[56]**)**.** The fact that the same power law is observed in many different organisms, suggests a universality of vascular architecture. Some explanations of this universality (such as the model in **[56]**) have appeared in the literature.

C.S. has verified this scaling law for human fetuses, using the placental weight as a proxy for the basic metabolic rate (**[57]**). This results in an allosteric scaling equation

$$log \text{ [placental weight]} = log \ \alpha \ + \ \beta \cdot log \text{ [birth weight]}, \qquad (*)$$

in which $\beta$ represents the unknown power. The calculated value of $\beta$ was $0.78 \pm 0.02$ in an excellent agreement with the rule.

Our DLA model is premised on the shape of the placenta reflecting an underlying vascular fractal. It is thus reasonable to expect, that a deviation from the normal round shape of a placenta with a central insertion of the umbilical cord will be correlated with abnormal placental vascular architecture. The latter should result in a deviation from the normal value of $\beta \approx 3/4$.

To test this prediction, we have employed several measures of deviation from a round placental shape. Firstly, we have calculated the standard deviation of the radius, measured from the point of insertion of the umbilical cord. For a round placenta with a central insertion, this value is *0*. However, this measure will not differentiate between a regular ellipse and a star. We have used another simple measure, the *roughness*, calculated as the perimeter of the shape divided by the perimeter of the smallest convex hull that contained the shape. It is equal to *1* for any convex shape, such as a circle. Descriptive statistics determined the percentage with values for both parameters within *10%* of the values for a round circle (*0, 1* respectively), and for the subsequent deciles of deviation.

In the table below, the results of Spearman's rank correlations are presented for our dataset. The variable *OutSdRCrd* is the deviation of the radius from the point of insertion, *OutRough* is the roughness. Both of them are significantly correlated with the variable *beta3_4*, which is the difference between the computed value of $\beta$ from the scaling equation (*), and the "ideal" value *0.75*. The inverse (-) association of the radial deviation and *beta3_4* means that as the deviation increases the difference between β observed and *0.75* is greater.

Our interpretation is confirmed by the lack of correlation of the variable *OutSdR* (the standard deviation of the radius calculated from the centroid of the placental shape) with the deviation from *0.75*. This is as we predict, since the centroid of the placental shape is an arbitrary geometrical center that does not directly relate to the underlying vascular architecture that ramifies out from the umbilical cord insertion site.

|  |  | *beta3_4* |
|---|---|---|
| *OutSdR* | Pearson Correlation | .020 |
|  | Significance | .485 |
|  | Number of measurements | 1199 |
| *OutSdRCrd* | Pearson Correlation | **-.076** |
|  | Significance | **.009** |
|  | Number of measurements | **1187** |
| *OutRough* | Pearson Correlation | **.091** |
|  | Significance | **.002** |
|  | Number of measurements | **1199** |

**Table 1:** *correlation of the deviation from a round shape with a deviation from the ¾ rule. Lower value of significance level means stronger evidence (e.g. 0.009 means 9 in a 1000 chance that the correlation happens by a coincidence).*

**Conclusion.**

We have presented a mechanism which accounts for two of the most common patterns of abnormal placental shapes (multilobate placentas and star-shaped placentas), specifically, changes in the arborization of the vascular tree. To confirm it, we have developed a dynamic model of growth for the vascular tree based on a DLA random growth process. We have demonstrated that the observed variability of shapes is explained by a change of the arborization parameter of the model at a single time instance. Thus, the time in gestation of determination of these two patterns may be distinct.

In addition, we have empirical evidence that the shape of the placenta does reflect the underlying vascular fractal; deviations from symmetric fractal expansion out from the umbilical cord insertion are associated with reduced placental functional efficiency, i.e., a smaller birth weight for the given placental weight.

The use of a fractal model of this type is reasonable to model the placenta. Its structure is little more than a thin tissue sheath covering a complex arborizing vascular network. Our empiric data are consistent with its function being intimately dependent on its structure, suggesting that our model will be germane to any studies of birth weight as a measure of the adequacy of the intrauterine environment. Restriction of placental vascular outgrowth is a potent model of fetal growth restrictions **[51]**. Placental growth within the uterine lining has been likened to tumor growth for decades (**[52]** , **[53]**). Fractals have likewise proved fruitful to analysis of retinal vasculature architecture **[54]** as well as the vascular structure of the heart and brain **[55]**.

Our model, that correlates placental structure to birth weight which has been used as a proxy for the adequacy of the intrauterine environment in most studies of the fetal origins of adult health risks (**[24]-[46]**) may be helpful in studies of the relationship of intrauterine stressors to the fetal modifications that are believed to underlie these lifelong health risks.

**Computer software.**

Numerical simulations of vascular trees were carried out using *dla-3d-placenta*, a Unix-based, ANSI C, 3-dimensional, diffusion-limited aggregation simulation package. In its development, we have used Mark Stock's *dla-nd* arbitrary-dimensional diffusion-limited aggregation simulator, a free software developed under the terms of the GNU General Public License as published by Free Software Foundation. For DLA cluster visualization we have used *PovRay*: a freeware ray tracing program available for a variety of computer platforms.

**Random number generators.**

Random number generators used in this study include:

- drand48(), an ANSI C system-supplied double-precision linear congruential uniform number generator, (generates sequence of integers $I_1$, $I_2$, $I_3$,..., each between 0 and $M$ by the recurrence relation $I_{n+1} = aI_n + c \pmod{M}$).
- ran1(long *idum), uniform random number generator of Park and Miller with Bays-Durham shuffle and added safeguards **[47]**.
- ran2(long *idum), long period random number generator of L'Ecuyer with Bays-Durham shuffle and added safeguards **[47]**.

**Affiliation of the authors:**
- M.Y. Department of Mathematics, University of Toronto, 40 St. George Street, Toronto, ON, Canada, M5S2E4
- C.S. Department of Psychiatry, New York University School of Medicine. 550 First Avenue, NY NY 10016, Department of Obstetrics and Gynecology, St Luke's Roosevelt Hospital, NY, NY, 10019.
- O.S. Department of Mechanical and Industrial Engineering, University of Toronto, 5 King's College Road, Toronto, ON, Canada M5S 3G8
- D.H. Department of Pathology, St Luke's Roosevelt Hospital, NY, NY, 10019.
- B.E. and J.T. Department of Obstetrics and Gynecology, University of North Carolina at Chapel Hill